\documentclass[pre]{revtex4}
\usepackage{graphicx}
\usepackage{dcolumn}
\usepackage{bm}

\begin{document}

\title{SPONTANEOUS AND STIMULATED SYNCHROTRON RADIATION FROM RELATIVISTIC 
ELECTRONS IN ION CHANNEL }

\author{I.~Kostyukov}
 \affiliation{Institute of Applied Physics, Russian Academy of Science, 
46 Uljanov St. 603950 Nizhny Novgorod, Russia}

\author{A.~Pukhov}
\affiliation{Institut fur Theoretische Physik I, Heinrich-Heine-Universitat 
Dusseldorf, 40225 Dusseldorf, Germany}

\date{\today} 

\begin{abstract}
Spontaneous and stimulated emission of an electron in the ion 
channel is studied. The emission processes are studied in the 
regime of high harmonic generation when the parameter of plasma 
wiggler strength is large. Like for conventional free electron 
laser, a synchrotron-like broadband spectrum is generated in 
this regime. The asymptotic expression for the radiation
spectrum of the spontaneous emission is derived. The radiation 
spectrum emitted from axisymmetric monoenergetic electron beam 
is analyzed. The gain of ion-channel synchrotron-radiation 
laser is calculated. Use of laser-produced ion channel for  
efficient X-ray generation is discussed.
\end{abstract}


\maketitle

\section{Introduction}

An electron dynamics in plasma-focusing channel has important 
applications to new plasma technologies, such as advanced 
accelerators \cite{esarey-review}, novel radiation sources, 
new types of lens \cite{lens}. It is a key phenomenon for 
ion-channel laser (ICL) \cite{ion channel laser},
ion-ripple laser \cite{ion ripple laser}, plasma-wiggler free electron
laser(FEL) \cite{plasma fel}, which are perspective candidates for high
brightness X-ray radiation sources. Such radiation sources are strongly
needed for experimental research in physics, chemistry, biology and in
engineering \cite{light sources}.

Resent experiments that explore the interaction of intense 
28.5-GeV electron beam with plasma at Stanford Linear Accelerator 
Center (SLAC) \cite{wang,joshi-review} have shown that ion 
channel can be successfully used to produce broadband X-ray radiation. 
Moreover, the high density of the ions in the channel provides much 
higher wiggler strength than that provided by a conventional magnet 
wiggler. This leads to a more effective generation of X-ray 
radiation than in conventional light sources and could be used for the
development of next generation of radiation sources.

To create ion channel, electron beam have to interact with plasma in
blow-out regime \cite{blow-out regime} when the electron beam density, 
$n_{b}$, is higher than plasma density, $n_{p}$. In this regime 
the electron beam charge quasistatically expels slow plasma electrons 
inside and around the beam volume and the ion channel is formed. 
Note that relativistic electrons of the beam are not expelled 
from the channel because of relativistic compensation of the beam 
electron charge force by the self-magnetic force \cite{book-beam}. 
The channel radius,  $r_{i}\simeq r_{0}\sqrt{n_{b}/n_{p}}$ is much 
more than electron beam radius, $r_{0}$, for dense $n_{b}\gg n_{p}$
and narrow $k_{p}r_{0}\ll 1$ electron beam \cite{beam radius}, where 
$k_{p}=c/\omega _{p}$ is the plasma skin depth, $\omega _{p}=\left( 4\pi
n_{p}e^{2}/m\right) ^{1/2}$ is the plasma frequency, $e$ is the electron
charge, $m$ is the electron mass and $c$ is the speed of light. If all
plasma electrons are expelled from the channel then the restoring force on
the beam electrons due to the ion charge is given by Gauss's low and in
cylindrical geometry is:
\begin{equation}
{\bf F}_{res}=m\omega _{p}^{2}{\bf r}_{\perp }/2,  
\label{restoring force}
\end{equation}
where ${\bf r}_{\perp }$ is the vector from an electron to the 
channel axis. The beam electrons\ in the ion channel will undergo 
betatron oscillations caused by this force. The wavelength for small
betatron oscillations is close to $\lambda _{b}=2\pi /k_{b}\simeq 
2\pi \sqrt{2\gamma }/k_{p}$, where $\gamma $ is the relativistic 
factor of the electron beam \cite{betatron oscillations}. 

\begin{figure}
   \begin{center}
   \begin{tabular}{c}
   \includegraphics[height=5cm,clip]{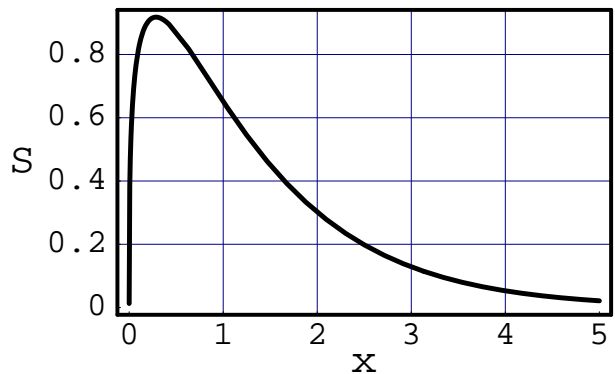}
   \end{tabular}
   \end{center}
   \caption[example] 
   { \label{synchrotron function} 
{\small The synchrotron radiation function $S(x)$ versus $x$.}
}
   \end{figure}

It is well known that accelerated charges emit electromagnetic (EM)
radiation \cite{jackson}. Therefore the electrons undergoing betatron
oscillations in ion channel will emit short-wavelength EM\ radiation. Some
features of this radiation spectrum have been studied in 
Ref.~ \cite{wang,joshi-review,esarey1}. The wavelength 
of the radiation is close to $\lambda \simeq \lambda _{b}/
\left( 2\gamma ^{2}\right) $ for small-amplitude
near-axis betatron oscillations. If the amplitude of the
betatron oscillations becomes large then electron radiates high harmonics.
If the plasma wiggler strength, 
\begin{equation}
K=\gamma k_{b}r_{0}=1.33\cdot 10^{-10}\sqrt{\gamma n_{e}
\left[ cm^{-3}\right]}r_{0}\left[ \mu m\right] ,  
\label{wiggler strength}
\end{equation}
is so high that $K\gg 1$ then radiation spectrum becomes quasi-continuous
broadband, where $r_{0}$ is amplitude of electron betatron oscillation in
the channel. The frequency dependence of the radiation spectrum becomes
similar to the one of synchrotron radiation spectrum which is determined by
the universal function 
\begin{equation}
S\left( \omega /\omega _{c}\right) =\left( \omega /\omega _{c}
\right) \int_{\omega /\omega _{c}}^{\infty }K_{5/3}\left(
x\right) dx,
\end{equation}
where $\omega _{c}$ is the critical frequency 
(see Fig.~\ref{synchrotron function}) \cite{jackson}. 
For frequencies well below the critical frequency $\left( \omega
\ll \omega _{c}\right) $ the spectrum increases with frequency as $\omega
^{2/3}$, reaches a maximum at$\ \sim 0.29\omega _{c}$, and then drops
exponentially to zero above $\omega _{c}$. The critical frequency for a
relativistic electron in an ion channel is \cite{esarey1} 
\begin{equation}
\omega _{c}=\frac{3}{2}\gamma ^{3}cr_{0}k_{b}^{2}=5.2\cdot 10^{-24}\gamma
^{2}n_{e}\left[ cm^{-3}\right] r_{0}\left[ \mu m\right] keV.
\label{crytical frequency}
\end{equation}%
Because of the strongly relativistic motion of the electron the 
emitted radiation is confined in very narrow angle $\theta \simeq 
K/\gamma $. Synchrotron radiation in ion channel has been observed in recent 
experiments \cite{wang}.

The averaged total power radiated by an electron undergoing betatron
oscillations is \cite{esarey1}
\begin{equation}
\left\langle P_{total}\right\rangle \simeq 
\frac{e^2 c}{12}N_{b} \gamma^{2} k_b^4 r_{0}^{2},  
\label{total power}
\end{equation}
where $N_{b}$ is the number of the betatron oscillations performed by 
the electron. We can introduce also the stopping power of an electron, 
that is the energy loss of an electron per unit distance 
\begin{equation}
Q=\left\langle P_{total}\right\rangle /c
\simeq 1.5\cdot 10^{-45}\left( \gamma n_{e}\left[ cm^{-3}\right] 
r_{0}\left[\mu m\right] \right) ^{2}\frac{MeV}{cm}.  
\label{stopping power}
\end{equation}
The averaged number of the photons with averaged energy 
$\hbar \omega _{c}$ is 
\begin{equation}
\left\langle N_{ph}\right\rangle \simeq \frac{4\pi }{9}
\frac{e^{2}}{\hbar c}N_{b}K\simeq 1.02\cdot 10^{-2}N_{b}K.  
\label{Photon number}
\end{equation}
It follows from Eq.~(\ref{total power}) that the radiated power 
is proportional to the squared density of ions in the channel. 
This fact has been confirmed in the experiments \cite{wang}. 
As it was mentioned above the ion density in the channel 
have to be less than electron density in the beam. This leads to 
the serious limits on the gain in radiated power. One of the ways 
to overcome the limits is to use laser-produced ion channel. 
The ion density in such channel can be about $10^{19}cm^{-3}$ 
\cite{laser channel,pukhov1} that is in 5 order higher than ion 
density in the channel produced in beam - plasma interaction 
experiments \cite{wang}. So use of ion channel produced by 
laser pulse could increase the power of X-ray radiation in 
$10^{10}$ times! More detailed discussion on use of relativistic 
laser channel for X-ray generation will be presented in
Conclusion.

Spontaneous emission in an ion channel has been studied in 
detail in Ref.~\cite{esarey1}. The general expression for the 
spectrum has been derived. It is a complex expression that 
involves the sum of products of the Bessel function. Numerical 
evaluation of the spectrum becomes difficult in the limit 
$K \gg 1$. The simple asymptotic expression for the angular 
dependence of the radiated spectrum has been derived for this 
limit only for directions that are perpendicular to the plane of 
the betatron oscillation. The one of the purposes of our paper is to
calculated the total angular dependence of the radiated spectrum.

Resonance interaction between EM radiation and the betatron 
oscillations of electron beam in ion channel leads to the bunching of 
the electron beam and then to the amplification (or damping) of 
EM radiation. It is a stimulated emission (or absorption). Stimulated 
emission is a basic process in ICLs \cite{ion channel laser} and FELs 
\cite{book-fel}. Reverse process - stimulated absorption leads to the
direct laser acceleration \cite{pukhov DLA} and to the magnetic field
generation \cite{kostyukov IFE} in relativistic laser channel. 
Unfortunately stimulated emission (absorption) in ion channel has not 
yet been explored in the limit $K\gg 1$ that is another objective of 
our paper.

The paper is organized as follows. In Section II the motion of an 
electron is studied and Hamiltonian formulation of the problem is 
presented. In Section III, the spontaneous emission from electrons 
undergoing betatron oscillations in ion channel is analyzed. The general 
asymptotic expression of the radiation spectrum for arbitrary angular 
dependence is derived for $K\gg 1$. The spectrum averaged over azimuthal 
angle is calculated for axially symmetrical electron beam in the limit 
$K\gg 1$. Section IV discusses the stimulated emission processes. The 
gain of ion - channel synchrotron - radiation laser is derived. A summary
discussion is presented in Sec. V.

\section{Electron dynamics in ion channel}

Relativistic equation of electron motion in cylindrical ion channel is
\begin{equation}
\frac{d{\bf p}}{dt}={\bf F}_{res},  \label{eq - betatron motion}
\end{equation}%
where ${\bf F}_{res}$ is the restoring force defined by Eq.~(\ref{restoring
force}). It follows from Eq.~(\ref{eq - betatron motion}) that momentum
along channel axis $p_{z}$ is a constant of motion. First we will consider
the radial betatron oscillation as it takes place when the center of an
electron bunch moves along the channel axis. Assuming that $p_{y}=0$ we get
the equation for $x$ -coordinate:
\begin{equation}
\frac{d^{2}x}{dt^{2}}=\frac{\gamma _{z}^{2}}{2\gamma ^{3}}x,
\label{motion - x}
\end{equation}
where we introduce the constant of motion $\gamma _{z}^{2}=1+p_{z}^{2}$ and
use the dimensionless units, normalizing the time to $\omega _{p}^{-1}$, the
length to $c/\omega _{p}$, the momentum to $mc$.

As Hamiltonian does not depend on time it is another constant of motion
\begin{equation}
H=\gamma +\frac{x^{2}}{4}=const=\gamma _{z}+\frac{r_{0}^{2}}{4},
\label{second constant}
\end{equation}
where $r_{0}$ is the amplitude of the betatron oscillation. We can express 
$\gamma $ as function of $x$ from the obtained relation to resolve 
Eq.~(\ref{motion - x}). Then transversal motion of electron can be reduced to the
oscillations in effective potential
\begin{equation}
U\left( x\right) =\frac{8\gamma _{z}^{2}}{\left( 4\gamma
_{z}+r_{0}^{2}-x^{2}\right) ^{2}}.  \label{effective potential}
\end{equation}%
The oscillations can be described in implicit form as
follows 
\begin{eqnarray}
t &=&r_{0}\frac{\sqrt{\nu ^{2}+4}}{\nu }E\left( arc\sin 
\left( \frac{x}{r_{0}}\right) ,\frac{\nu ^{2}}{\nu ^{2}+4}\right)   
\nonumber \\
&&-\frac{2}{\nu \sqrt{\nu ^{2}+4}}F\left( arc\sin 
\left( \frac{x}{r_{0}}\right) ,\frac{\nu ^{2}}{\nu ^{2}+4}\right) ,  
\label{betatron orbit}
\end{eqnarray}
where $E(x,k)$ and $F(x,k)$ are the Elliptic integrals of the first and the
second kinds \cite{special function}, respectively, and $\nu $ is
\begin{equation}
\nu ^{2}=r_{0}^{2}/2\gamma _{z}.  \label{nu definition}
\end{equation}
The period of betatron oscillations is
\begin{equation}
T_{b}=\frac{2\pi }{\omega _{b}}=4r_{0}\frac{\sqrt{\nu ^{2}+4}}{\nu }\left[
E\left( \frac{\nu ^{2}}{\nu ^{2}+4}\right) -\frac{2}{\nu ^{2}+4}K\left( 
\frac{\nu ^{2}}{\nu ^{2}+4}\right) \right] ,  
\label{betatron frequency}
\end{equation}%
where $K(x)$ and $E(x)$ are the complete Elliptic integrals of the first and
the second kinds \cite{special function}, respectively. In the limit $\nu
^{2}\ll 1$, parameter $\nu $ is the ratio between the longitudinal and
transversal energy of the electron  
\begin{equation}
\nu ^{2}\simeq p_{x}^{2}/\gamma _{z}^{2}\simeq p_{\perp }^{2}/p_{z}^{2},
\label{nu}
\end{equation}%
where $p_{x}$ is the maximum of the transversal momentum which is at the
channel axis $x=0$. In most applications transversal moment of the electron 
is much less than longitudinal one, so we assume that $\nu \ll 1$. Then we 
can use expansion in $\nu $ to describe betatron oscillations:
\begin{equation}
\omega _{b}=\frac{1}{\sqrt{2\gamma _{z}}}\left( 1-\frac{3}{8}\nu
^{2}+...\right) ,  \label{betatron expansion}
\end{equation}
\begin{equation}
x\left( t\right) =r_{0}\sin \left( \omega _{b}t\right) -r_{0}\frac{3}{64}\nu
^{2}\sin \left( 3\omega _{b}t\right) +....  \label{betatron amplitude}
\end{equation}%
Using Eq.~(\ref{second constant})\ we obtain the relations for the electron
orbit in the zeroth order in $\nu $ that coincides with ones calculated in
Ref.~\cite{esarey1}
\begin{equation}
\omega _{b}\simeq \frac{1}{\sqrt{2\gamma _{z}}},  \label{wb}
\end{equation}
\begin{equation}
x\left( t\right) \simeq r_{0}\sin \left( \omega _{b}t\right) ,  \label{x(t)}
\end{equation}
\begin{equation}
y\left( t\right) \simeq 0,  \label{y(t)}
\end{equation}
\begin{equation}
z\left( t\right) \simeq z_{0}+\frac{p_{z}}{\gamma _{z}}\left( 1-\frac{\nu
^{2}}{4}\right) t-r_{0}\frac{p_{z}}{\gamma _{z}}\frac{\nu }{8}\sin \left(
2\omega _{b}t\right) .  \label{z(t)}
\end{equation}%
Notice that parameter $\nu $ coincides with the expression $k_{b}r_{0}$
introduced in Ref.~\cite{esarey1} and plasma wiggler strength parameter can
be expressed through $\nu $ as $K\simeq \gamma _{z}\nu \simeq p_{\perp }$.

More general regime of the betatron motion when $p_y \neq 0$ and an
electron orbit is not plane has been considered and classified in 
Ref.~\cite{kostyukov IFE} in the limit $p_{z}\gg p_{\perp }$. Equations of 
motion take a form in this case
\begin{equation}
x\left( t\right) \simeq \frac{p_{x}}{\sqrt{2\gamma _{z}}}\sin \left( \omega
_{b}t\right) ,  \label{x1}
\end{equation}
\begin{equation}
y\left( t\right) \simeq \frac{p_{y}}{\sqrt{2\gamma _{z}}}\sin \left( \omega
_{b}t+\psi \right) ,  \label{y1}
\end{equation}
\begin{equation}
z\left( t\right) \simeq z_{0}+\frac{p_{z}}{\gamma _{z}}\left[ 1-\frac{%
p_{x}^{2}+p_{y}^{2}}{4\gamma _{z}^{2}}-\frac{p_{x}^{2}}{8\gamma _{z}^{2}}%
\sin \left( 2\omega _{b}t\right) +\frac{p_{y}^{2}}{8\gamma _{z}^{2}}\sin
\left( 2\omega _{b}t+2\psi \right) \right] ,  \label{z1}
\end{equation}
where $\psi $ is the phase difference between oscillations along $x$-axis and 
$y$-axis, $p_{x}$ and $p_{y}$ are the maximum of the electron momentum along 
$x$-axis and along $y$-axis, respectively, that occurs at the channel axis 
($x=0$, $y=0$). If angular momentum of the electron, $L=p_{y}x-p_{x}y$ is
equal to zero, then electron executes radial harmonic oscillations through
the origin with amplitude $r_{0}=2\sqrt{H}$. If $L=\pm L_{\max }=\pm
H/\omega _{b}$ then an electron performs circular motion with radius $r_{0}$. 
In the general case (an arbitrary value of $-L_{\max }<L<L_{\max }$) the
electron trajectory is an ellipse-like and is confined between the maximal
radius, $\sqrt{\left( H+\sqrt{H^2-\omega _{b}^{2}L^{2}}\right)}$,
and minimal radius, $\sqrt{\left( H-\sqrt{H^2-\omega _{b}^{2}L^{2}}\right)}$.

\section{Spontaneous emission}

Using Eqs.~(\ref{x1}) - (\ref{z1}) for electron trajectory, the energy
spectrum radiated by an electron can be calculated \cite{jackson,esarey1}.
The total radiation flux can be separated in two independent components with
polarization in the ${\bf e}_{\theta }$ and ${\bf e}_{\phi }$ directions,
where unit vectors ${\bf e}_{\theta }$ and ${\bf e}_{\phi }$ correspond to
spherical coordinate system: $x=r\sin \theta \cos \phi $, $y=r\sin \theta
\sin \phi $, $z=r\cos \theta $. Then energy radiated per frequency $\omega $
per solid angle in the direction ${\bf k}=\left( \omega /c\right) \left( 
{\bf e}_{x}\sin \theta \cos \phi +{\bf e}_{y}\sin \theta \sin \phi +{\bf e}_{z}
\cos \theta \right) $ during the interaction time $T$ is \cite{esarey1}: 
\begin{equation}
\frac{d W_{spon,j}}{d \omega  d\Omega }=\frac{e^{2}}{4\pi ^{2}c}\eta ^{2}
\left\vert I_{j}\right\vert ^{2},
\label{Q defenition}
\end{equation}
\begin{equation}
I_{\theta }=\int\limits_{-T/2}^{T/2}\left( \frac{dx}{dt}\cos \theta \cos
\phi +\frac{dy}{dt}\cos \theta \sin \phi -\frac{dz}{dt}\sin \theta \right)
e^{i\Psi }dt,  \label{I Theta}
\end{equation}
\begin{equation}
I_{\phi }=\int\limits_{-T/2}^{T/2}\left( \frac{dx}{dt}\sin \phi -\frac{dy}{
dt}\cos \phi \right) e^{i\Psi }dt,  \label{I Fi}
\end{equation}
\begin{equation}
\Psi =\eta \left[ t-x\left( t\right) \sin \theta \cos \phi -y\left( t\right)
\sin \theta \sin \phi -z\left( t\right) \cos \theta \right] ,  
\label{phasa}
\end{equation}%
where $j=\theta ,\phi $ is the polarization index, the electron trajectory
is given by Eqs.~(\ref{x1}) - (\ref{z1}) and $\eta =\omega /\omega _{p}$.
The final results can be expressed as double infinite series of the Bessel
function products (see Eqs.~(32) - (41) in Ref.~\cite{esarey1} or can be
expressed by the infinite series of the generalized Bessel function
introduced in quantum electrodynamics \cite{general Bessel}). Unfortunately
the series converge slowly in the limit $K\gg 1$ that makes the numerical
evaluation of the radiation spectrum difficult. Energy spectrum and angular
dependence of radiation have been derived only  for directions that are
perpendicular to the plane of the betatron oscillation (i. e., for $\phi
=\pi /2$) \cite{esarey1}. To extend this result we will use saddle point
method \cite{saddle point} to evaluate integrals (\ref{I Theta}) 
and (\ref{I Fi}).

   \begin{figure}
   \begin{center}
   \begin{tabular}{c}
   \includegraphics[height=6cm,clip]{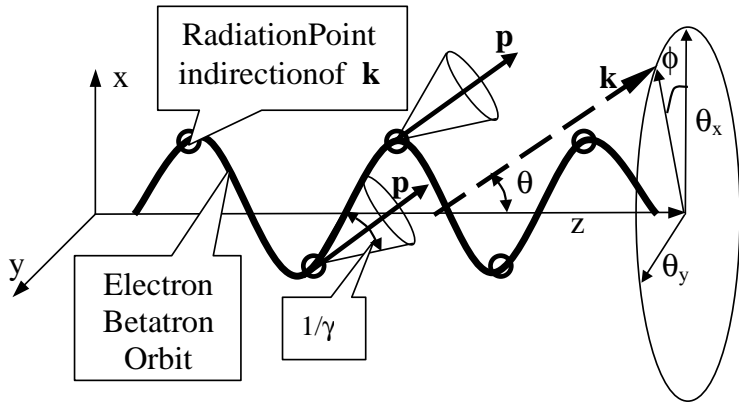}
   \end{tabular}
   \end{center}
   \caption[example] 
   { \label{saddle points} 
{\small Schematic of synchrotron radiation in an ion channel. Open
circles show the points on the electron trajectory when the electron emits
in the direction of ${\bf k}$.}
}
   \end{figure} 

It is well known that radiation of accelerated relativistic electron is
beamed in a very narrow cone in the direction of the electron momentum
vector, and is seen by the observer as a short pulse of radiation as the
searchlight beam sweeps across the observation point \cite{jackson} (see
Fig.~\ref{saddle points}). So the time moments when the electron momentum, 
${\bf p}$, is directed along wave number, ${\bf k}$, give the main 
contribution in integrals ( \ref{I Theta}) and (\ref{I Fi}). As the 
pulse duration is very short it is necessary to know the electron 
momentum and the electron position over only a small arc of the 
trajectory whose tangent points in the direction that is close to the 
direction of ${\bf k}$. Then we can expand integrand in 
Eqs.~(\ref{I Theta}) and (\ref{I Fi}) about this time moments and 
perform integration. This approach implies that we approximate the part
of the electron trajectory near these time moments by the arc of a 
circular path \cite{jackson}. In this case the radiation will have 
well-known synchrotron-like spectrum. It is noted in Ref.~\cite{landau} 
that another necessary (the first is that the electron have to be 
relativistic) condition for use of the synchrotron radiation approach is 
that the electron deflection angle should be much more than simultaneous 
angle spread to which radiation is emitted. The electron momentum 
oscillates in cone angle $\sim p_{\perp}/p_{z}$ (electron deflection 
angle). The radiation of the relativistic particle is confined to 
angle $1/\gamma $ \cite{jackson}. So the validity condition for 
synchrotron radiation approach is $p_{\perp }/p_{z}\gg 1/\gamma $ that 
is equivalent to the condition for high harmonics generation discussed 
in Introduction $K\gg 1$. Therefore the radiation spectrum can be
approximated by the synchrotron one if
\begin{equation}
\gamma \gg p_{\perp }\gg 1.  
\label{condition synchrotron}
\end{equation}
This condition is easily satisfied in experiments. 

The arguments presented above justify the use of the saddle point method 
\cite{saddle point} to evaluate integrals (\ref{I Theta}) and (\ref{I Fi}).
We can expand phase $\Psi $ about the moment of time $\xi _{n}= 
\omega_{b}t_{n}$ to the third order:
\begin{equation}
\Psi =\Psi _{0}+b_{1}(\xi -\xi _{n})+b_{2}(\xi -\xi _{n})^{2}+b_{3}
(\xi -\xi_{n})^{3}, 
\label{phase expnation}
\end{equation}
\begin{eqnarray}
\Psi _{0} &=&\Psi \left( \xi _{n}\right) =\alpha _{0}\xi _{n}
-\alpha_{x}\sin \left( \xi _{n}\right) -\alpha _{y}\sin \left( \xi _{n}
+\psi \right)
\nonumber \\
&&+\alpha _{zx}\sin \left( 2\xi _{n}\right) +\alpha _{zy}\sin 
\left( 2\xi_{n}+2\psi \right) ,  
\label{phase 0}
\end{eqnarray}
\begin{eqnarray}
b_{1} &=&\left. \frac{d\Psi }{d\xi }\right\vert _{\xi =
\xi _{n}}=\alpha_{0}-\alpha _{x}\cos \left( \xi _{n}\right) 
-\alpha _{y}\cos \left( \xi_{n}+\psi \right)  
\nonumber \\&&
+2\alpha _{zx}\cos \left( 2\xi _{n}\right) +2\alpha _{zy}\cos 
\left( 2\xi_{n}+2\psi \right) ,  
\label{phase 1}
\end{eqnarray}
\begin{eqnarray}
b_{2} &=&\left. \frac{d^{2}\Psi }{d\xi ^{2}}\right\vert _{\xi =\xi
_{n}}=\alpha _{x}\sin \left( \xi _{n}\right) +\alpha _{y}\sin 
\left( \xi_{n}+\psi \right)  
\nonumber \\ &&
-4\alpha _{zx}\sin \left( 2\xi _{n}\right) -4\alpha _{zy}\sin 
\left( 2\xi_{n}+2\psi \right) ,  
\label{phase 2}
\end{eqnarray}
\begin{eqnarray}
b_{3} &=&\left. \frac{d^{3}\Psi }{d\xi ^{3}}\right\vert _{\xi =
\xi_{n}}=\alpha _{x}\cos \left( \xi _{n}\right) +\alpha _{y}\cos 
\left( \xi_{n}+\psi \right)  
\nonumber \\&&
-8\alpha _{zx}\cos \left( 2\xi _{n}\right) -8\alpha _{zy}\cos 
\left( 2\xi_{n}+2\psi \right) ,  
\label{phase 3}
\end{eqnarray}
where
\begin{equation}
\alpha _{0}=\frac{\eta }{\omega _{b}}\left[ 1-\frac{p_{z}}
{\gamma _{z}}\left( 1-\frac{p_{x}^{2}+p_{y}^{2}}{4\gamma _{z}^{2}}
\sin \theta \cos \phi \right) \right] ,  
\label{alpha 0}
\end{equation}
\begin{equation}
\alpha _{x}=\frac{\eta }{\omega _{b}}\frac{p_{x}}{\gamma _{z}}
\sin \theta \cos \phi ,  
\label{alpha x}
\end{equation}
\begin{equation}
\alpha _{y}=\frac{\eta }{\omega _{b}}\frac{p_{y}}{\gamma _{z}}
\sin \theta \cos \phi ,  \label{alpha y}
\end{equation}
\begin{equation}
\alpha _{zx}=\frac{\eta }{\omega _{b}}\frac{p_{z}}
{\gamma _{z}}\frac{p_{x}^{2}}{8\gamma _{z}^{2}}\cos \theta ,  
\label{alpha zx}
\end{equation}%
\begin{equation}
\alpha _{zy}=\frac{\eta }{\omega _{b}}\frac{p_{z}}{\gamma _{z}}
\frac{p_{y}^{2}}{8\gamma _{z}^{2}}\cos \theta .  
\label{alpha zy}
\end{equation}%
Then we can expand the pre-exponent factors in Eqs.~(\ref{I Theta}) 
and (\ref{I Fi}) about the moment of time $\xi _{n}=\omega _{b}t_{n}$ 
o the first order: 
\begin{equation}
\frac{dx}{dt}\cos \theta \cos \phi +\frac{dy}{dt}\cos \theta \sin 
\phi -\frac{dz}{dt}\sin \theta =B_{\theta ,n}+D_{\theta ,n}(\xi 
-\xi _{n}),
\label{Factor Theta}
\end{equation}%
\begin{equation}
\frac{dx}{dt}\sin \phi -\frac{dy}{dt}\cos \phi =B_{\phi ,n}+D_{\phi ,n}
(\xi -\xi _{n}),  
\label{Factor Fi}
\end{equation}%
where 
\begin{eqnarray}
B_{\theta ,n} &=&\left( \frac{p_{x}}{\gamma _{z}}\cos \xi _{n}\cos \phi +%
\frac{p_{y}}{\gamma _{z}}\cos \left( \xi _{n}+\psi \right) \sin \phi \right)
\cos \theta   \nonumber \\
&&-\sin \theta \frac{p_{z}}{\gamma _{z}}\left[ 1-\frac{p_{x}^{2}}{4\gamma
_{z}^{2}}\cos \left( 2\xi _{n}\right) -\frac{p_{y}^{2}}{4\gamma _{z}^{2}}%
\cos \left( 2\xi _{n}+2\psi \right) \right] ,  \label{B Theta}
\end{eqnarray}%
\begin{eqnarray}
D_{\theta ,n} &=&-\left[ \frac{p_{x}}{\gamma _{z}}\sin \xi _{n}\cos \phi +%
\frac{p_{y}}{\gamma _{z}}\sin \left( \xi _{n}+\psi \right) \sin \phi \right]
\cos \theta   \nonumber \\
&&-\frac{p_{z}}{\gamma _{z}}\frac{p_{x}^{2}}{2\gamma _{z}^{2}}\sin \theta
\sin \left( 2\xi _{n}\right) -\frac{p_{z}}{\gamma _{z}}\frac{p_{y}^{2}}{%
2\gamma _{z}^{2}}\sin \theta \sin \left( 2\xi _{n}+2\psi \right) ,
\label{D Theta}
\end{eqnarray}%
\begin{equation}
B_{\phi ,n}=\frac{p_{x}}{\gamma _{z}}\sin \phi \cos \left( \xi _{n}\right) -%
\frac{p_{y}}{\gamma _{z}}\cos \phi \cos \left( \xi _{n}+\psi \right) ,
\label{B Fi}
\end{equation}%
\begin{equation}
D_{\phi ,n}=-\frac{p_{x}}{\gamma _{z}}\sin \phi \sin \left( \xi _{n}\right) +%
\frac{p_{y}}{\gamma _{z}}\cos \phi \sin \left( \xi _{n}+\psi \right) .
\label{D Fi}
\end{equation}%
Notice that it is sufficient to keep the leading term in the pre-exponent
factor while only terms that are much less than unity can be neglected in
the exponent argument.

The main contribution to the integral comes from the neighborhood 
of the saddle points specified by $d\Psi /d\xi =0$ \cite{saddle point}. 
The first-order term  in phase expansion~(\ref{phase expnation}) can be 
written as follows 
\begin{equation}
\left. \frac{d\Psi }{d\xi }\right\vert _{\xi =\xi _{n}}=\frac{\eta }
{\omega_{b}}\left( 1-\frac{{\bf k\cdot p}}{k\gamma }\right) 
\simeq \frac{\eta }{2\omega _{b}}\left( \frac{1}{\gamma ^{2}}
+\varphi ^{2}\right) ,
\label{first derivative}
\end{equation}
where $\varphi $ is the angle between ${\bf k}$ and ${\bf p}$. It follows
from Eq.~(\ref{first derivative}) that $d\Psi /d\xi $ is minimal and close
to zero at $\varphi =0$ when the electron momentum is directed along ${\bf k}
$ that agrees with the qualitative argument presented above.
 
For simplicity we assume that $p_{y}=0$ that is the electron orbit is plane
and the betatron oscillations is radial. It follows from 
Eq.~(\ref{phase 1}) that the values of $\xi $ whose
neighborhood gives the main contribution to the integral are defined by the
relation
\begin{equation}
\cos \xi _{n}=\frac{\gamma _{z}}{p_{x}}\tan \theta \cos \phi .
\label{saddle point}
\end{equation}%
It is seen from Eq.~(\ref{saddle point}) and Fig.~3 that the number of
saddle points is $2N_{b}=\omega _{b}T/\pi $ that is the number of times when
the direction of the electron momentum and the direction of the wave number
coincides. It can be shown \cite{jackson} that the second-order term can be 
neglected in Eq.~(\ref{phase expnation}). Then Eqs.(\ref{I Theta}) and 
(\ref{I Fi}) take a form: 
\begin{eqnarray}
I_{j} &=&\frac{1}{\omega _{b}}\sum_{n=1}^{2N_{b}}\exp \left( i\Psi
_{0,n}\right) R_{j,n},  \label{I-definition} \\
R_{j,n} &=&\int_{-\infty }^{+\infty }ds_{n}\left(
B_{j,n}+D_{j,n}s_{n}\right) \exp \left[ ib_{1,n}s_{n}+ib_{3,n}s_{n}^{3}%
\right] ,  \label{R}
\end{eqnarray}%
where $j=\theta ,\phi $ is the polarization index, 
$s_{n}=\xi _{n}-\xi _{0}$, $\Psi _{0,n}$ is the value 
of phase $\Psi $ in the $n$-th saddle point.  

It follows from the definitions of saddle point 
Eq.~(\ref{saddle point}) and
from Eqs.~(\ref{phase expnation}) - (\ref{D Fi}) that $
R_{j,n}=R_{j,n-1}=R_{j}$, $B_{j,n}=B_{j,n-1}=B_{j}$, $
D_{j,n}=D_{j,n-1}=D_{j} $, $b_{m,n}=b_{m,n-1}=b_{m}$. Performing 
integration in Eq.~(\ref{R}) we obtain 
\begin{equation}
R_{j}=\sqrt{\frac{8b_{1}}{3b_{3}}}\left[ B_{j}K_{1/3}\left( \sqrt{\frac{%
8b_{1}^{3}}{9b_{3}}}\right) +D_{j}\sqrt{\frac{2b_{1}}{b_{3}}}K_{2/3}\left( 
\sqrt{\frac{8b_{1}^{3}}{9b_{3}}}\right) \right] ,  \label{R result}
\end{equation}%
where $K_{1/3}(x)$ and $K_{2/3}(x)$ are the modified Bessel function 
\cite{special function}. In the synchrotron regime of radiation  
$\omega \sim \omega _{c}\gg \omega _{p}$ and $\Psi _{0,n}\gg 1$ 
then we can write for large number of the betatron periods 
($N_{b}\gg 1$ )
\begin{equation}
\left\vert \sum_{n=1}^{2N_{b}}\exp \left( i\Psi _{0,n}\right) \right\vert
^{2}\simeq 2N_{b}.  \label{Nb}
\end{equation}

Using Eqs.~(\ref{nu}), (\ref{Q defenition}), (\ref{R result}) and condition 
(\ref{condition synchrotron}) we finally get 
\begin{equation}
\frac{d W_{spon,\theta }}{d \omega  d\Omega } =2N_{b}\frac{e^{2}\eta ^{2}
\rho ^{2}\chi }{3\pi ^{2}c}\left[\sin \theta \sin ^{2}\phi K_{1/3}(q)+
\sqrt{\chi }\cos \phi K_{2/3}(q)\right]^{2},  
\label{Power spontaneous - theta}
\end{equation}
\begin{equation}
\frac{d W_{spon,\phi }}{d \omega  d\Omega }=2N_{b}\frac{e^{2}\eta ^{2}
\rho ^{2}\chi \sin ^{2}\phi }{3\pi^{2}c}\left[ \sin \theta \cos \phi 
K_{1/3}(q)-\sqrt{\chi }K_{2/3}(q)\right]^{2},  
\label{Power spontaneous - fi}
\end{equation}
where 
\begin{equation}
\rho =\sqrt{\frac{2\gamma _{z}}{p_{x}^{2}/\gamma _{z}^{2}-\sin ^{2}\theta
\cos ^{2}\phi }}  \label{curvature radius}
\end{equation}%
is the curvature radius of the circular path that is used to approximate the
part of the electron trajectory where electron emits in the direction of 
${\bf k}$, 
\begin{equation}
\chi =\gamma _{z}^{-2}+\sin ^{2}\theta \sin ^{2}\phi ,
\label{angle argument}
\end{equation}
\begin{equation}
q=\frac{1}{3}\eta \rho \chi ^{3/2}.  \label{Bessel-argument}
\end{equation}%
The total radiation of the spontaneous emission from the electron in the
channel is 
\begin{eqnarray}
\frac{d W_{spon}}{d \omega  d\Omega }&=&\frac{d W_{spon,\theta }}
{d \omega  d\Omega }+\frac{d W_{spon,\phi }}{d \omega  d\Omega } 
\nonumber \\
&=&2N_{b}\frac{e^{2}\eta ^{2}\rho ^{2}\chi ^{2}}{3\pi ^{2}c}
\left[ \frac{\sin ^{2}\theta \sin ^{2}\phi }{\chi }K_{1/3}^{2}
(q)+K_{2/3}^{2}(q)\right] .
\label{Power spontaneous - total}
\end{eqnarray}
This is the general expression for the angular distribution of the radiation
emitted by an a relativistic electron in ion channel and this is the one of
the main results of the paper.

Let us consider some limiting cases. It follows from condition 
(\ref{condition synchrotron}) that $\theta \ll 1$. Then in the 
limit $\phi =\pi /2 $ Eq.~(\ref{Power spontaneous - total}) takes a form 
\begin{equation}
\frac{d W_{spon}}{d \omega  d\Omega }=N_{b}\frac{6e^{2}}{\pi ^{2}c}
\frac{\gamma _{z}^{2}q^{2}}{\left(1+\gamma _{z}^{2}\theta ^{2}\right) }
\left[ \frac{\gamma _{z}^{2}\theta ^{2}}{\left( 1+\gamma _{z}^{2}
\theta ^{2}\right) }K_{1/3}^{2}(q)+K_{2/3}^{2}(q)\right] ,  
\label{esarey formula}
\end{equation}
that coincides with the asymptotic spectrum emitted by the relativistic
electron in the channel for $\phi =\pi /2$ (see Eq.~(64) in 
Ref.~\cite{esarey1}). It was discussed above that in the limit 
$K\gg 1$ the radiation emitted from an electron at the given moment 
of time is similar to the synchrotron radiation emitted from 
an electron in instantaneously circular motion with the same curvature 
radius. Really, introducing notation $\varphi =\sin \theta \sin \phi $ 
and using relation $\gamma \simeq \gamma _{z}\simeq p_{z}$ we can 
reduce Eq.~(\ref{Power spontaneous - total}) to the known form
\begin{equation}
\frac{d W_{spon}}{d \omega  d\Omega }=2N_{b}\frac{e^{2}\left( \eta \rho 
\right) ^{2}}{3\pi ^{2}c} \left(\frac{1}{\gamma ^{2}}+\varphi ^{2}
\right) ^{2}\left[ K_{2/3}^{2}(q)+\frac{\varphi ^{2}}{1/\gamma ^{2}
+\varphi ^{2}}K_{1/3}^{2}(q)\right] ,
\label{jackson formula}
\end{equation}
that coincides with the expression for energy radiated by a relativistic
electron in instantaneously circular motion with radius $\rho $ per unit
frequency per unit solid angle $2\pi \cos \varphi d\varphi $ after $2N_{b}$
revolutions (see Eq.~(14.83) in Ref.~\cite{jackson}). 

\begin{figure}
   \begin{center}
   \begin{tabular}{c}
   \includegraphics[height=5cm,clip]{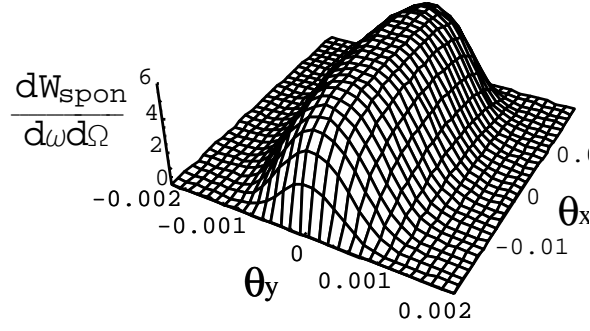}
   \end{tabular}
   \end{center}
   \caption[example] 
   { \label{spontaneous} 
{\small Angular distribution of the spontaneous emission spectrum 
$\frac{d W_{spon}}{d \omega  d\Omega } $ (arbitrary units) versus angles 
$\theta _{x}$ and $\theta _{y}$ from a single electron with $\gamma_{z}=1000$, 
$K=p_{x}=20$ for $\omega =0.5\omega _{c}$.}
}
   \end{figure}

To visualize our results we use new angle coordinates $\theta _{x}=\sin
\theta \cos \phi $, $\theta _{y}=\sin \theta \sin \phi $ instead of the
spherical one $(\theta ,\varphi )$ (see Fig.~\ref{saddle points} 
and Fig.~\ref{spontaneous}). It is seen  from Fig.~\ref{spontaneous} 
that there is no radiation for $\theta _{x}>\theta _{\max }$
because the argument of the Bessel function in 
Eq.~(\ref{Power spontaneous - total}) goes to infinity 
and the Bessel function goes to zero for $\theta _{x}=\theta _{\max }$. 
Therefore in our approximation the emission angle along $x$-axis is 
confined to the electron deflection angle $\theta _{\max }=p_{x}/p_{z}$. 
However it is evident that electrons
with maximal deflection angle radiation $p_{x}/p_{z}$ emit radiation up to
the angle $\theta =\pm \left( \theta _{\max }+\gamma ^{-1}\right) $. The
emission angle is confined to the angle $\sim 1/\gamma $ in the direction of 
$y$-axis which is normal to the electron orbit plane. Hence our results
agree with the qualitative analysis in Ref.~\cite{esarey1}.

Averaging Eq.~(\ref{Power spontaneous - total}) over $\theta _{y}$ we obtain 
\begin{equation}
\int_{-1}^{1}\frac{d W_{spon}}{d \omega  d\Omega } d\theta _{y}
=\frac{\sqrt{3}}{2}N_{b}\frac{e^{2}\gamma_{z}}{\pi c}
S\left( \frac{2}{3}\frac{\sqrt{2}\eta }{\gamma _{z}^{3/2}
\sqrt{p_{x}^{2}-\gamma _{z}^{2}\theta _{x}^{2}}}\right) ,  
\label{averaging over y}
\end{equation}
where $S\left( x\right) $ is the universal function mentioned in
Introduction. Expression (\ref{averaging over y}) can be considered as 
a radiation power from electron flow in which electrons are evenly 
distributed over $y$-axis (over direction which is normal to the 
betatron oscillation plane).

Let us now consider a monoenergetic, axisymmetric electron beam. Radiation
spectrum from electron beam with electron distribution function $f\left( 
{\bf p}\right) $ is defined by the relation:
\begin{equation}
\left\langle \frac{d W_{spon}}{d \omega  d\Omega }\right\rangle 
=\int_{0}^{2\pi }\frac{d\phi }{2\pi } \int_{-\infty }^{+\infty }d{\bf p}
f\left( {\bf p}\right) \frac{d W_{spon}}{d \omega  d\Omega } 
\left( {\bf k,p}\right) .  
\label{beam radiation}
\end{equation}
Let all electrons have the same longitudinal and transversal energy before 
interaction and the electron distribution function is
\begin{equation}
f\left( {\bf p}\right) =\delta (p_{z}-\gamma _{z})\delta 
(\sqrt{p_{x}^{2}+p_{y}^{2}}-p_{\bot }).  
\label{edf}
\end{equation}
Then the radiation spectrum from the beam takes a form
\begin{equation}
\left\langle \frac{d W_{spon}}{d \omega  d\Omega } \right\rangle 
=\int_{0}^{2\pi }\frac{d\phi }{2\pi }\frac{d W_{spon}}{d \omega  d\Omega }.  
\label{beam spontaneous}
\end{equation}
In the limit of near-axis radiation $\theta \ll 1/\gamma _{z} $ the 
radiation spectrum is
\begin{equation}
\left\langle \frac{d W_{spon}}{d \omega  d\Omega } \right\rangle \simeq 
2N_{b}\frac{3e^{2}}{\pi ^{2}c} \left(\frac{\sqrt{2}\eta \gamma _{z}^{3}}
{3p_{x}}\right) ^{2}K_{2/3}^{2} \left( \frac{\sqrt{2}\eta \gamma _{z}^{3}}
{3p_{x}}\right) .
\label{Q spont averaged q1 << 1}
\end{equation}%
In the reverse limit $\theta \gg 1/\gamma _{z} $ the main contribution to
Eq.~(\ref{beam spontaneous}) is given by the small values of $\phi \ll 1$
that correspond to the minimum of $q$. The radiation spectrum in this limit
is 
\begin{equation}
\left\langle \frac{d W_{spon}}{d \omega  d\Omega } \right\rangle \simeq 
\frac{\sqrt{3}}{2}N_{b} \frac{e^{2}\gamma_{z}}{\pi c\theta }S\left( \frac{2}
{3}\frac{\sqrt{2}\eta }{\gamma _{z}^{3/2}\sqrt{p_{x}^{2}-\gamma _{z}^{2}
\theta ^{2}}}\right).
\label{Q spont averaged q1 >>1}
\end{equation}
Note that the obtained expression coincides with 
Eq.~(\ref{averaging over y}). 

\section{Stimulated synchrotron radiation in ion channel}

When EM wave of approximately the same frequency as the 
spontaneous emission is driven through the ion channel 
simultaneously with the electron beam, a significant exchange of 
energy between the beam and the wave can occur and
can lead to the coherent efficient amplification of the wave energy.
This amplification can be explained in terms of the stimulated 
emission processes which is determined the operation of ICL. 
Unlike ICL theory \cite{ion channel laser}, here we will consider 
regime of strong wiggler ($K \gg 1$) when the emission process 
is close to synchrotron one. This is the regime  of 
ion-channel synchrotron-radiation laser (ICSRL). 

The difference between spontaneous and stimulated emissions is the
following. The radiation fields generated by an electron undergoing betatron
oscillations has a phase which depends on time of arrival of the electrons
at the channel entrance. The fields produced by different electrons in a
uniform input beam have a random phase relation to each other and sum up
incoherently. This lead to incoherent spontaneous radiation. Contrary to the
spontaneous emission, electrons in ion channel can be driven by an external
EM wave in synchronous oscillation, the phase of which is no
longer random but locked to the phase of the wave. As a result the external
wave causes the bunching of the electron beam and more efficient interaction
between the beam and the wave. Then the radiation fields of different
electrons sum up coherently to each other and to the external wave, leading
either to a decrease or to an increase of the power, depending on whether
destructive or constructive interference is realized. So the external wave
is either absorbed or amplified. In quantum approach the
amplification/absorption can be viewed as a transition, forced by the
external wave, between the quantum states of the electron in 
the ion channel with photon absorption/emission. An amplification process of
this kind is called a stimulated emission. It is known from laser physics
that stimulated emission can be much more efficient and powerful than 
spontaneous one.

As known in quantum physics \cite{haitler} there is the relation between the
spontaneous and stimulated emission. Using this fact the elementary quantum
methods based on the Einstein's coefficients have been used to study the
instability of EM waves in cosmic plasmas \cite{astro}. Particularly, the
synchrotron instability in a cold magnetoactive plasmas has been identified 
\cite{synhrotron instability}. The relation between spontaneous and
stimulated emission of the electrons in undulators is called Madey's theorem
in the theory of FELs \cite{Madey's theorem}. It can be considered as 
extension of Einstein's coefficient method to the classical limit. 
Generalized Madey's theorem \cite{book-fel} enables us to reduce the problem 
of ICSRL gain to the problem solved in Sec. III that is the calculation 
of the power of spontaneous emission in ion channel.

To use Madey's theorem we should formulate the problem within Hamiltonian
approach. An electron motion in an ion channel and in EM wave 
can be described by a relativistic Hamiltonian
\begin{equation}
H=\sqrt{1+({\bf p}_{\perp }+{\bf A}_{i})^{2}+p_{z}^{2}}+\frac{x^{2}+y^{2}}
{4},  
\label{Hamiltonian in laser}
\end{equation}%
where ${\bf A}_{i}$ is the vector potential of the wave with $\theta $ or 
$\phi $ polarizations ($i=\theta ,\phi $): 
\begin{eqnarray}
\mathbf{A}_{\theta } &=&A_{0}\left( \mathbf{e}_{x}\cos \theta \cos 
\phi +\mathbf{e}_{y}\cos \theta \sin \phi -\mathbf{e}_{z}\sin 
\theta \right)  
\nonumber \\
&&\times \exp \left[ i\eta t-i\eta \left( \mathbf{k}\cdot 
\mathbf{r}\right) /k\right] ,  
\label{A Theta}
\end{eqnarray}%
\begin{equation}
\mathbf{A}_{\phi }=A_{0}\left( \mathbf{e}_{x}\sin \phi 
-\mathbf{e}_{y}\cos \phi \right) \exp \left[ i\eta t-i\eta 
\left( \mathbf{k}\cdot \mathbf{r}\right) /k\right] .  
\label{A Fi}
\end{equation}
Hamiltonian (\ref{Hamiltonian in laser}) is again written in the
dimensionless units, normalizing the time to $\omega _{p}^{-1}$, the length
to $c/\omega _{p}$, the momentum to $mc$, the vector potential to 
$mc^{2}/e$. As usual we assume that the time of
arrival of the electrons at the channel entrance is random. Assuming that
external EM wave is weak we can consider it as perturbation and
use the perturbation theory to calculate the work done upon the electron
beam by EM wave with $j$-polarization. Then it follows from the generalized
Madey's theorem (see Eq.~(5.52) in Ref.~\cite{book-fel}) that this work per
beam electron is
\begin{equation}
\left\langle W_{j} \right\rangle =\frac{1}{2}\frac{\partial
\left\langle \gamma _{1,j}^{2}\right\rangle }{\partial \gamma _{0}}
+\frac{1}{2}\frac{{\bf k}_{\perp }}{k}\cdot \frac{\partial \left\langle 
\gamma_{1,j}^{2}\right\rangle }{\partial {\bf p}_{\perp }},  
\label{Madey 1}
\end{equation}
where $\gamma _{1,j}$ is the first-order work done upon a single electron 
moving along the unperturbed electron trajectory ${\bf r}^0\left( t\right) $ 
by  EM wave with $j$-polarization 
\begin{equation}
\gamma _{1,j}=\int_{-\infty }^{+\infty }dt\frac{d{\bf r}^{0}
\left( t\right)}{dt}\cdot {\bf A}_{i} \left[ t,{\bf r}^0\left( t\right) 
\right] ,
\label{gamma 1}
\end{equation}
$\gamma _{0}$ and ${\bf p}_{\perp }$ are the electron energy and the
transversal momentum of the electron before the interaction. The unperturbed
electron trajectory ${\bf r}^0(t)$ is determined by 
Eqs.~(\ref{x1}) - (\ref{z1}). Averaging in Eq.~(\ref{Madey 1}) 
implies the averaging over the time of arrival of the
electrons at the channel entrance. Mathematical statement of the Madey's 
theorem is that the second-order quantity, 
$\left\langle W_{j}\right\rangle $, is
proportional to the average squared first-order quantities, $\gamma _{1,j}$.
Therefore Madey's theorem essentially simplifies calculations in the
framework of the perturbation theory. Using instead of variables $\gamma _{0}
$ and ${\bf p}_{\perp }$ the initial value of the electron momentum 
${\bf p}=({\bf p}_{\perp },p_{z})$, Eq.~(\ref{Madey 1}) can be 
also rewritten in more symmetric form 
\begin{equation}
\left\langle W_{j}\right\rangle =\frac{1}{2}\frac{{\bf k}}{k}\cdot 
\frac{\partial \left\langle \gamma _{1,j}^{2}\right\rangle }
{\partial {\bf p}}.  
\label{Madey 2}
\end{equation}
Similar to the FEL theory we introduce the incremental gain of ICSRL as a
ratio between a power generated by the electron beam and the incoming EM
wave power 
\begin{equation}
\Gamma _{j}=\frac{2\lambda ^{2}r_{e}n_{b}\left\langle W_{j} \right\rangle }
{\pi A_{0}^{2}},  
\label{gain}
\end{equation}
where $n_{b}$ is the density of the electron beam, $r_{e}=mc^{2}/e^{2}$ is
the classical electron radius, $\lambda =2\pi c/\omega $ is wavelength of EM
wave. It should be noted that to calculate 
$\left\langle W_{j}\right\rangle $ we
consider the given EM wave and do not consider the dynamics of EM wave
during the interaction. Hence our calculations is valid for $\Gamma \ll 1$.
This regime of interaction is called in the FEL theory as a small-signal
small-gain regime \cite{book-fel}.

It follows from Eqs.~(\ref{I Theta}), (\ref{I Fi}) and (\ref{gamma 1}) that 
$\gamma _{1,j}$ is proportional to $I_{j}$. That is the particular case of
the general reciprocity relation between the far field of a moving electron
and the work done by a plane EM wave on it \cite{book-fel}. Therefore we can
express quantity $\left\langle \gamma _{1,j}^{2}\right\rangle $ through the
energy of the spontaneous emission energy radiated per frequency per solid
angle  
\begin{equation}
\left\langle \gamma _{1,j}^{2}\right\rangle =A_{0}^{2}\left( \frac{e^{2}}
{4\pi ^{2}c}\frac{\omega ^{2}}{\omega _{p}^{2}}\right) ^{-1}
\frac{d W_{spon,j}}{d \omega  d\Omega }.
\label{gamma 1 and Q spon}
\end{equation}
and can express the gain in term of $\frac{d W_{spon,j}}{d \omega  d\Omega }$
\begin{equation}
\Gamma _{j}=\frac{\lambda ^{2}r_{e}n_{e}}{\pi }\left( \frac{e^{2}}{4\pi
^{2}c }\frac{\omega ^{2}}{\omega _{p}^{2}}\right) ^{-1}\left( \frac{{\bf k}}{%
k} \cdot \frac{\partial }{\partial {\bf p}}\right) 
\frac{d W_{spon,j}}{d \omega  d\Omega }.
\label{spontaneous and stimulayed}
\end{equation}

It has been noted in previous Section that efficient spontaneous emission 
in direction ${\bf k}$ takes place only at a short moment of time when 
the electron moment is directed along ${\bf k}$. Therefore we can conclude 
from Eq.~(\ref{spontaneous and stimulayed}) that interaction with EM
wave propagating in direction of ${\bf k}$ is only possible at the same
moments of time. It was mentioned in Sec. III (see Fig.~3) that the 
number of the interactions moments is $2N_{b}$. The number of the electron 
oscillations in EM wave between interaction moments is extremely large: 
$ N_{EMW}\simeq \omega /\omega _b \simeq \omega _c /\omega_b \simeq 
3\gamma ^{2}K /2$.  
For example for SLAC experiments $N_{EMW}\simeq 9.2\cdot 10^{9}$. 
So we can consider the electron phases as random before each 
interaction moments and, therefore, can consider each interaction 
moments independently. Using this
fact Eq.~(\ref{spontaneous and stimulayed}) can be rewritten as follows 
\begin{equation}
\Gamma _{j}=2N_{b}\frac{\lambda ^{2}r_{e}n_{e}}{\pi }
\left( \frac{e^{2}} {4\pi^{2}c}\frac{\omega ^{2}}{\omega _{p}^{2}}
\right) ^{-1}\left( \frac{{\bf k}}{k}\cdot \frac{\partial }
{\partial {\bf p}}\right) 
\left( \frac{d W_{spon,j}}{d \omega  d\Omega } \frac{1}{N_b}\right) .  
\label{spontaneous and stimulayed -2}
\end{equation}

For simplicity we assume that $p_{y}=0$. To take derivatives in 
Eq.~(\ref{spontaneous and stimulayed -2}) we should present 
$\frac{d W_{spon,j}}{d \omega  d\Omega }$ as a
function of momentum, ${\bf p}$, and use the simplifying assumptions 
$p_{y}=0 $ only after performing differentiation in 
Eq.~(\ref{spontaneous and stimulayed -2}). This is because 
the electron motion in the ion channel and
in EM wave is no longer plane. Although the unperturbed betatron
oscillations is in the plane $y=0$ the action of EM wave leads 
to the small oscillations along $y$-axis in the first-order 
of perturbation theory.

\begin{figure}
   \begin{center}
   \begin{tabular}{c}
   \includegraphics[height=10cm,clip]{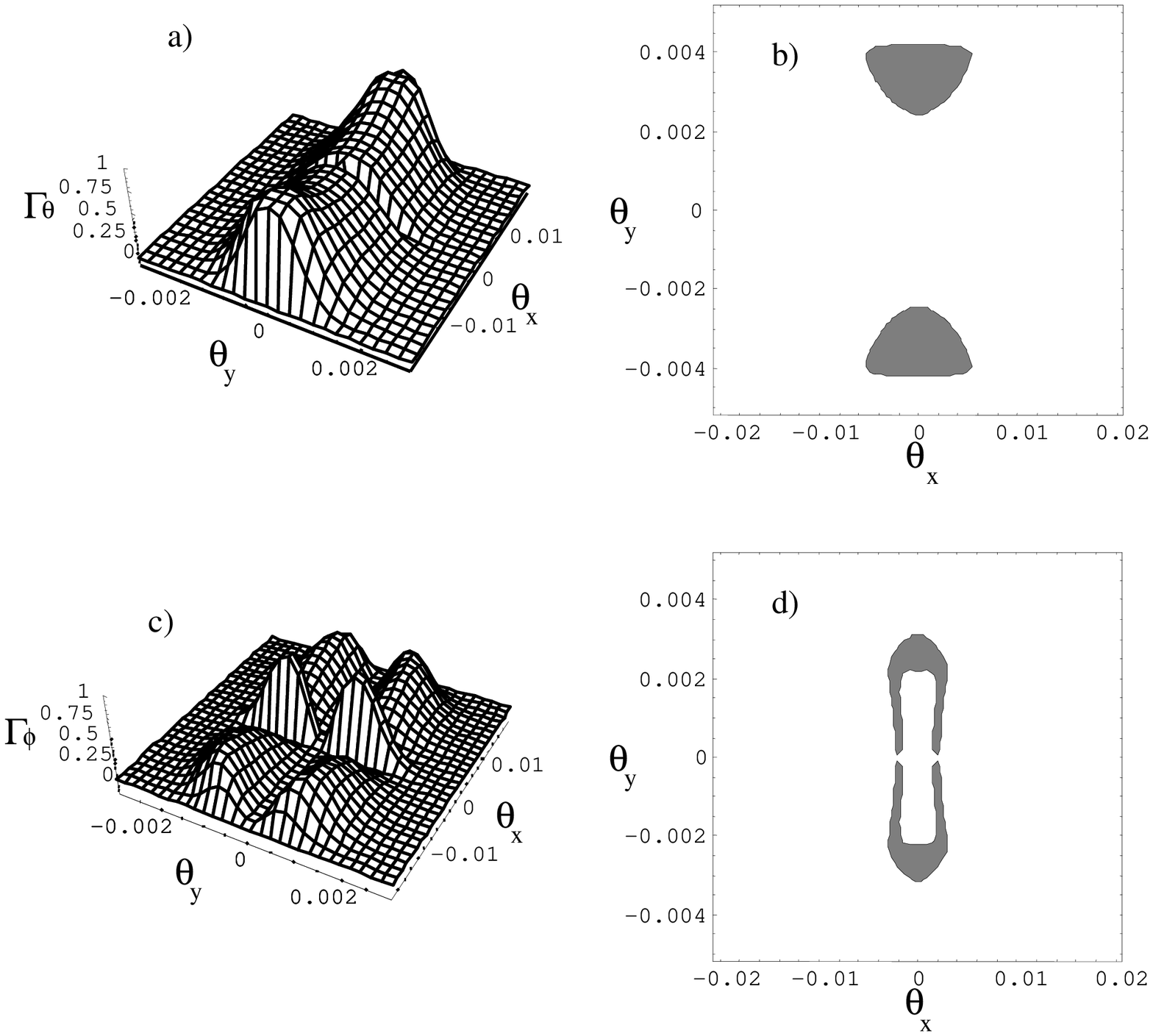}
   \end{tabular}
   \end{center}
   \caption[example] 
   { \label{icsrl} 
{\small 

{\bf a)} $\Gamma _{\theta }$ (arbitrary units), for 
$\theta $-polarized EM wave with 
$\omega =\omega _{c}$ versus angles $\theta _{x}$
and $\theta _{y}$ for electrons with $\gamma _{z}=500$, $K=p_{x}=10$.

{\bf b)} The domains of the angles $\theta _{x}$ and $\theta _{y}$ 
where $\theta $-polarized EM wave with $\omega =\omega _{c}$ is 
amplified by the electrons $\Gamma _{\theta } < 0$, grey region) and 
where the EM wave is absorbed by the electrons
($\Gamma _{\theta } > 0$, white region) for $\gamma _{z}=500$, 
$K=p_{x}=10$. The angles are given in radians.

{\bf c)} $\Gamma _{\phi }$ (arbitrary units), for 
$\phi $-polarized EM wave with $\omega =\omega _{c}$ versus angles 
$\theta _{x}$ and $\theta _{y}$ for electrons with $\gamma _{z}=500$, 
$K=p_{x}=10$.

{\bf d)} The domains of the angles $\theta _{x}$ and $\theta _{y}$ 
where $\phi $-polarized EM wave with $\omega =\omega _{c}$ 
is amplified by the electrons ($\Gamma _{\phi } < 0$, grey region) 
and where the EM wave is absorbed by the electrons ($\Gamma _{\phi } 
>0$, white region) for $\gamma _{z}=500$, $K=p_{x}=10$. The angles
 are given in radians.
 
}
}
   \end{figure}

It follows from Eqs.~(\ref{Q defenition}), (\ref{I-definition}) 
and (\ref{Nb}) that 
\begin{equation}
\left( \frac{e^{2}}{4\pi ^{2}c}\frac{\omega ^{2}}{\omega _{p}^{2}}
\right)^{-1}\frac{d W_{spon,j}}{d \omega  d\Omega } \frac{1}{N_b}
=\frac{2}{\omega _{b}^{2}}\eta ^{2}\left\vert R_{j}\right\vert ^{2}.  
\label{Q and R}
\end{equation}
Then using Eqs.~(\ref{phase 1})-(\ref{R result}) we can perform
differentiation in Eq.~(\ref{spontaneous and stimulayed}). To do it we have
to take into account that $\gamma _{z}$, $\omega _{b}$ are the function of 
$p_{z}$: $\gamma _{z}\left( p_{z}\right) =\sqrt{1+p_{z}^{2}}$, $\omega
_{b}\left( p_{z}\right) =$ $1/\sqrt{2\gamma _{z}\left( p_{z}\right) }$.
After performing differentiation we can put $p_{y}=0$. Then simplifying 
the obtained expression with help of
MATHEMATICA \cite{MATHEMATICA} the gain of ICSRL can be derived
\begin{eqnarray}
\Gamma _{\phi } &=& N_{b}\frac{\sqrt{2}\lambda ^{2}r_{e}n_{e}}{9\pi }\frac{
\rho ^{4}\sin ^{2}\phi }{\gamma _{z}^{3}p_{x}}\left[ \sin \theta \cos \phi
K_{1/3}(q)-\sqrt{\chi }K_{2/3}(q)\right]  \label{Power stim - fi} \\
&&\times \left[ \sqrt{\chi }\left( 3qT_{0}+\sin \theta \sqrt{\chi }
T_{1}\right) K_{1/3}(q)\right.  \nonumber \\
&&-\left. \left( 3q\sin \theta \cos \phi T_{0}+\chi ^{3/2}T_{2}\right)
K_{2/3}(q)\right] ,  \nonumber \\
T_{0} &=&\sin ^{2}\theta \cos ^{2}\phi \left[ 3p_{x}^{3}-8p_{x}^{2}\gamma
_{z}\sin \theta \cos \phi -p_{x}\gamma _{z}^{2}\sin ^{2}\theta \cos ^{2}\phi
\right.  \nonumber \\
&&\left. +2\gamma _{z}^{3}\sin ^{3}\theta \cos \phi \left( 1+2\cos ^{2}\phi
\right) \right] -3p_{x}^{3},  \nonumber \\
T_{1} &=&6p_{x}^{3}\cos \phi +2p_{x}^{2}\gamma _{z}\sin \theta \left(
3-8\cos ^{2}\phi \right)  \nonumber \\
&&-2p_{x}\gamma _{z}^{2}\sin ^{2}\theta \cos ^{3}\phi -2\gamma _{z}^{3}\sin
^{3}\theta \cos ^{2}\phi (1-4\cos ^{2}\phi ),  \nonumber \\
T_{2} &=&3p_{x}^{3}-8p_{x}^{2}\gamma _{z}\sin \theta \cos \phi -p_{x}\gamma
_{z}^{2}\sin ^{2}\theta \cos ^{2}\phi  \nonumber \\
&&+2\gamma _{z}^{3}\sin ^{3}\theta \cos \phi (1+2\cos ^{2}\phi ),  \nonumber
\end{eqnarray}
\begin{eqnarray}
\Gamma _{\theta } &=&N_{b}\frac{\sqrt{2}\lambda ^{2}r_{e}n_{e}}{9\pi }
\frac{\rho ^{4}}{\gamma _{z}^{3}p_{x}}\left[ \sin \theta \sin ^{2}\phi 
K_{1/3}(q)+\sqrt{\chi }\cos \phi K_{2/3}(q)\right]  
\label{Power stim - theta} \\
&&\times \left[ \sqrt{\chi }\left( 3q\cos \phi T_{0}+\sin \theta 
\sqrt{\chi }T_{3}\right) K_{1/3}(q)\right.  
\nonumber \\
&&\left. +\left( 3q\sin \theta \sin ^{2}\phi T_{0}
+\chi ^{3/2}T_{4}\right)K_{2/3}(q)\right],  
\nonumber \\
T_{3} &=&6p_{x}^{3}-16p_{x}^{2}\gamma _{z}\sin \theta \cos \phi
-2p_{x}\gamma _{z}^{2}\sin ^{2}\theta \cos ^{2}\phi  \nonumber \\
&&+4\gamma _{z}^{3}\sin ^{3}\theta \cos \phi \left( 1+2\cos ^{2}\phi \right)
,  \nonumber \\
T_{4} &=&3p_{x}^{3}\cos \phi +2p_{x}^{2}\gamma _{z}\sin \theta \left(
3-4\cos ^{2}\phi \right) +p_{x}\gamma _{z}^{2}\sin ^{2}\theta \cos ^{3}\phi 
\nonumber \\ &&-4\gamma _{z}^{3}\sin ^{3}\theta \cos ^{2}\phi \sin ^{2}\phi .  
\nonumber
\end{eqnarray}
We have also checked the obtained results by the numerical differentiation
for some values of parameters. EM wave can be amplified only if the wave 
propagates in some small angle to the axis $z$. It is seen from 
Fig.~\ref{icsrl} that there is no amplification of EM wave when the wave 
propagates along the channel axis.

\begin{figure}
   \begin{center}
   \begin{tabular}{c}
   \includegraphics[height=4cm,clip]{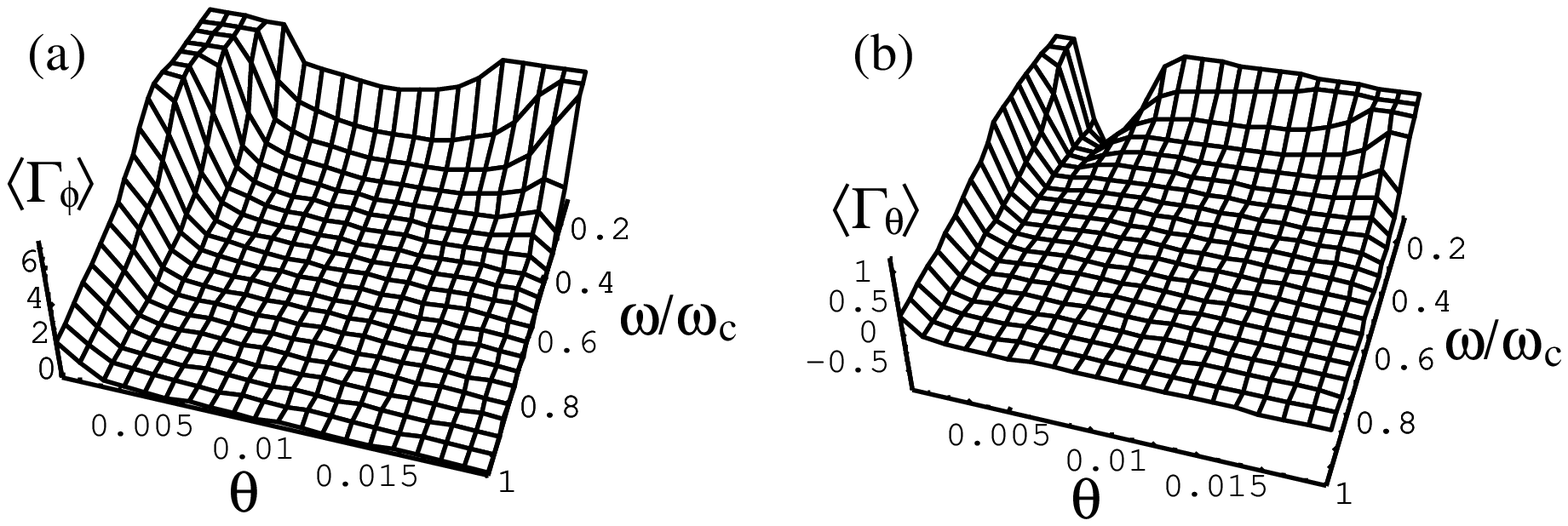}
   \end{tabular}
   \end{center}
   \caption[example] 
   { \label{icsrl averaged} 
{\small 

{\bf a)} $\left\langle \Gamma _{\phi }\right\rangle $
(arbitrary units) for axisymmetric electron beam with 
$\gamma _{z}=500$, $p_{\perp }=10$ and $\phi $-polarized 
EM wave versus angle $\theta $ and normolized frequency 
$\omega /\omega _{c}$.

{\bf b)} $\left\langle \Gamma _{\theta }\right\rangle $
(arbitrary units) for axisymmetric electron beam with 
$\gamma _{z}=500$, $p_{\perp }=10$ and $\theta $-polarized EM 
wave versus angle $\theta $ and normolized 
frequency $\omega /\omega _{c}$.
}
}
   \end{figure}

Eqs.~(\ref{Power spontaneous - fi}) and (\ref{Power spontaneous - theta})
have been derived under assumption that all electrons have the same momentum
before interaction. Now we will consider a monoenergetic, axisymmetric
electron beam with electron distribution function given by Eq.~(\ref{edf}).
ICSRL gain for such electron beam is defined by the relation
\begin{equation}
\left\langle \Gamma _{j}\right\rangle =\int_{0}^{2\pi }\frac{d\phi }{2\pi }%
\Gamma _{j}\left( {\bf p,}\phi \right) .  \label{gain averaged}
\end{equation}%
We have performed integration in Eq.~(\ref{gain averaged}) numerically. It
is seen from Fig.~\ref{icsrl averaged} that there is no amplification of 
EM wave by the beam $\phi $ for given parameters. We are not able to 
find the wave amplification by the axisymmetric electron beam at least 
for considered beam parameters.

\section{Discussion and Conclusions}

First we would like to discuss the use a laser-produced ion channel 
for X-ray generation. As it was mentioned in Introduction an ion 
channel in plasma can be produced by an electron beam itself.
However in this case the plasma density have to be less than beam 
density. Unfortunately the density of relativistic beam cannot be 
very high because of technology reasons. This leads to the serious 
limitation on the gain in radiated power as the power is quadratic
in plasma density. Use of high-power laser could overcome this 
limitation. High-power laser pulse can expel plasma electrons by 
ponderomotive force and create the ion channel behind the pulse 
\cite{laser channel}. Moreover, in strongly nonlinear "bubble" 
regime \cite{pukhov1} electrons are completely evacuated from the 
first half-plasma wave excited behind the laser pulse. The ion 
density in this "bubble" is higher in many order of magnitude than 
that in the ion channel formed in the beam - plasma interaction. 
For example, the ion density in the "bubble" can be as high 
as $10^{19}$cm$^{-3}$ \cite{pukhov1,laser channel} that is in 
$10^{5}$ times higher than that in the beam - plasma 
interaction experiments \cite {wang}. Therefore the radiated 
power in laser-produced channel can be in $10^{10}$ times 
higher than that in the experiments \cite {wang} for the 
same number of the betatron oscillations. 
The "bubble" moves with group velocity of the laser pulse which 
is close to the speed of light. Relativistic electron bunch 
injected into the "bubble" can propagate inside the "bubble" 
over a long distance. Hence, in spite of the small length
of the "bubble"  the electrons can oscillate in the "bubble" 
a lot of time.

   \begin{figure}
   \begin{center}
   \begin{tabular}{c}
   \includegraphics[height=6cm,clip]{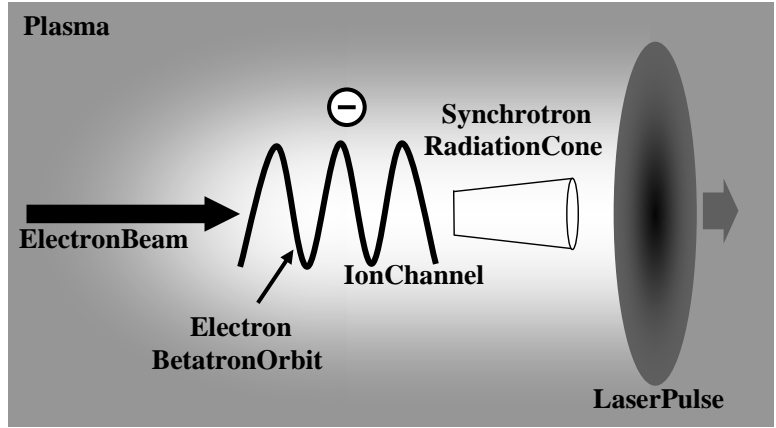}
   \end{tabular}
   \end{center}
   \caption[example] 
   { \label{synchrotron laser} 
{\small Schematic of the spontaneous emission from external
electron beam undergoing betatron oscillations in a 
laser-produced ion channel.}
}
   \end{figure} 

It should be noted that the self-generated forces of the electron bunch as
well as of the relativistic plasma electrons in the column can be neglected
because they cancel each other \cite{book-beam}. The
condition to neglect the self-generated forces is $n_{e}\ll n_{i}
\gamma ^{2}$ that is easily satisfied in modern experiments. 
The number of the betatron
oscillations performed by an electron can be calculated as a ratio of the
time during which electron passes through the "bubble" to the
period of the betatron oscillation: 
\begin{equation}
N_{b}=\frac{\omega _{p}L}{2\pi \sqrt{2\gamma }(c-v_{gr})},
\label{Nb in laser}
\end{equation}
where $v_{gr}$ is the group velocity of the laser pulse.

Some estimations for parameters in possible and carried out experiments are
presented in Table I. To calculate parameters in Table I we assume that 
$L\simeq 30\mu m$ according to the numerical simulations \cite{pukhov1}. The
group velocity of the laser pulse in rarefied plasma ($\omega_L \gg \omega_p$) 
can be estimated as $v_{gr}/c\simeq 1-\omega _{p}^{2}/\left( 2\omega _{L}^{2}
\right) $, where $\omega _{L}$ is the laser frequency \cite{pukhov1}. Then 
Eq.~(\ref{Nb in laser}) can be rewritten as follows 
\begin{equation}
N_{b}\simeq 4.7\cdot 10^{10}\frac{1}{\lambda _{L}\left[ \mu m\right] \sqrt{%
\gamma n_{e}\left[ cm^{-3}\right] }}\left( \frac{L}{\lambda _{L}}\right) ,
\label{Nb in laser - new}
\end{equation}%
where $\lambda _{L}$ is the laser wavelength.
\begin{table}[h]
\caption{Parameters of SLAC experiments \cite{joshi-review} (a); 
parameters of interaction between 30-GeV electron beam and 
laser-produced ion channel (b); parameters of interaction 
between 30-MeV electron beam generated in laser-plasma interaction 
\cite{pukhov1} and laser-produced ion channel (c). 
 }
\label{tab:fonts}
\begin{center}
\begin{tabular}{|c|c|c|c|c|c|c|c|c|}
\hline
\rule[-1ex]{0pt}{3.5ex} $$ & $r_0 $ & $\gamma $ & $n_i [cm^{-3}]$ & 
$K$ & $\omega_c [MeV]$ & $N_b$ & $Q [\frac{MeV}{cm}] $ & $N_{ph}$ 
\\ \hline \rule[-1ex]{0pt}{3.5ex} 
$a$ & $40$ & $6\cdot 10^4$ & $2\cdot 10^{14}$ & $17$ & 
$0.12$ & $1.5$ & $2.5\cdot 10^{-4}$ & $0.24$ 
\\ \hline \rule[-1ex]{0pt}{3.5ex} 
$b$ & $10$ & $6\cdot 10^{4}$ & $1\cdot 10^{19}$ & $1030$ & 
$1889$ & $2$ & $5.3\cdot 10^4 $ & $21 $ 
\\ \hline \rule[-1ex]{0pt}{3.5ex} 
$c$ & $10$ & $6\cdot 10^{2}$ & $1\cdot 10^{19}$ & $103$ & 
$0.76$ & $20$ & $5.3$ & $21 $ 
\\ \hline
\end{tabular}
\end{center}
\end{table}
It is seen from Table I that use of the laser-produced ion channel can
dramatically increase the power of X-ray emission. 

In this paper we have studied spontaneous and stimulated emission from
electrons undergoing betatron motion in ion channel. We calculate the period
of nonlinear betatron oscillations. The method based on the Bessel function
expansion is used in Ref.~\cite{esarey1} to calculate the spectrum of the
spontaneous emission in ion channel. However, in synchrotron regime of
emission ($K>>1$) the angular distribution of the radiation has been derived
in this paper only in the direction which is perpendicular to the electron
orbit plane. We extended this result to the arbitrary directions. The
generalized Madey's theorem was used to calculate the electron energy gain
of ICSRL. The calculation shows that there is no wave amplification when the
EM wave propagates along the channel axis and the amplification
takes place only when the wave propagates at some small angle to the channel
axis.

Our numerical simulations show that there is no amplification for
axisymmetric electron beam at least for considered parameters of the beam.
However further analysis is required to justify this conclusion. It may be
possible to overcome this difficulty by appropriately tailoring the electron
beam, that is, a narrow electron beam could be injected off-axis such that
all of the beam electrons execute approximately the same betatron orbit.

To calculate the radiation spectrum and the gain of ICSRL for electron beam,
we used a very simple distribution function. Further investigations should
include the more realistic electron distribution functions. The gain of
ICSRL was calculated in the small-signal small-gain limit. However, it is
more interesting for application is to explore large-gain regime of ICSRL
that needs further investigations.

\begin{acknowledgments}
One of the authors (I.~K.) gratefully acknowledges the hospitality of
Institute for Theoretical Physics of Duesseldorf University. This work was
support in part by Alexander von Humboldt Foundation (Germany) and by the
Russian Fund for Fundamental Research (Grants No 01-02-16575, No 01-02-06488
and by Russian Academy of Science (Grant N 1999-37).
\end{acknowledgments}

\end{document}